\documentclass[sigconf]{acmart}
\usepackage{array}
\usepackage{xcolor}

\usepackage{amsmath,amsfonts,bm}

\def\eqref#1{equation~\ref{#1}}

\def\1{\bm{1}}

\DeclareMathAlphabet{\mathsfit}{\encodingdefault}{\sfdefault}{m}{sl}
\SetMathAlphabet{\mathsfit}{bold}{\encodingdefault}{\sfdefault}{bx}{n}

\AtBeginDocument{%
  \providecommand\BibTeX{{%
    \normalfont B\kern-0.5em{\scshape i\kern-0.25em b}\kern-0.8em\TeX}}}

\begin{document}

\title{FNSPID: A Comprehensive Financial News Dataset in Time Series}

\author{Zihan Dong}
\email{zdong7@ncsu.edu}
\orcid{0003-4079-7520}
\authornotemark[1]
\affiliation{%
  \institution{North Carolina State University}
  \streetaddress{890 Oval Dr}
  \city{Raleigh}
  \state{North Carolina}
  \country{USA}
  \postcode{27607}
}

\author{Xinyu Fan}
\affiliation{%
  \institution{SiChuan University}
  \streetaddress{No.24, South Section 1, Yihuan Road}
  \city{SiChuan}
  \state{Sichuan}
  \country{China}}
\email{fanxinyu@stu.scu.edu.cn}

\author{Zhiyuan Peng}
\affiliation{%
  \institution{North Carolina State University}
  \streetaddress{890 Oval Dr}
  \city{Raleigh}
  \state{North Carolina}
  \country{USA}
  \postcode{27607}
}


\renewcommand{\shortauthors}{Dong and Fan, et al.}


\begin{abstract}
Financial market predictions utilize historical data to anticipate future stock prices and market trends. Traditionally, these predictions have focused on the statistical analysis of quantitative factors, such as stock prices, trading volumes, inflation rates, and changes in industrial production. Recent advancements in large language models motivate the integrated financial analysis of both sentiment data, particularly market news, and numerical factors. Nonetheless, this methodology frequently encounters constraints due to the paucity of extensive datasets that amalgamate both quantitative and qualitative sentiment analyses. To address this challenge, we introduce a large-scale financial dataset, namely, Financial News and Stock Price Integration Dataset (FNSPID). It comprises 29.7 million stock prices and 15.7 million time-aligned financial news records for 4,775 S\&P500 companies, covering the period from 1999 to 2023, sourced from 4 stock market news websites. We demonstrate that FNSPID excels existing stock market datasets in scale and diversity while uniquely incorporating sentiment information. Through financial analysis experiments on FNSPID, we propose: (1) the dataset's size and quality significantly boost market prediction accuracy; (2) adding sentiment scores modestly enhances performance on the transformer-based model; (3) a reproducible procedure that can update the dataset. Completed work, code, documentation, and examples are available at \url{github.com/Zdong104/FNSPID}. FNSPID offers unprecedented opportunities for the financial research community to advance predictive modeling and analysis.
\end{abstract}



\maketitle

\section{Introduction} 
For decades, time series regression models have been a cornerstone in developing financial valuation methods. This approach is pivotal not only in traditional finance models but also in artificial intelligence for financial forecasting, a field marked by the complexity and unpredictability of market patterns.

Traditional financial market analysis adopts the Fama-French Three-Factor Model (FFM)~\cite{fama1992cross}, and the Chen, Roll, and Ross Arbitrage Pricing Theory (APT)~\cite{chen1983APT} which are both pivotal in asset pricing. These models use linear regression to analyze returns but do not focus on specific market highs and lows. Both models' reliance on historical data limits their effectiveness in anticipating future market shifts or unprecedented events like financial crises.

Emerging machine learning (ML) techniques have shown promise in addressing these limitations. Previous studies demonstrate ML's effectiveness over traditional models~\cite{Sheth2023, Kurani2023}. Moreover, Billah and Bhuiyan highlighted the superiority of integrating stock price and news sentiments in deep learning (DL) techniques in stock market prediction~\cite{Billah2024}. These emerging methods, utilizing models like Long Short-Term Memory (LSTM), Recurrent Neural Network (RNN)~\cite{manurung2018algorithm}, and Reinforcement Learning (RL) methods, have demonstrated substantial improvement in timing market movements, a crucial aspect where traditional models fell short.~\cite{RLstockprediction2001Jae, data4030110, WU2020142, DeepLearning2021Yang}.

\begin{table*}[!t]
    \centering
    \small 
    \setlength\tabcolsep{4pt} 
    \begin{tabular}{*{10}{c}{>{\centering\arraybackslash}m{1.5cm}}}
        \toprule
        & \textbf{FNSPID (ours)} & Reuters & Benzinga & Bloomberg & Lenta & Lutz's & Farimani's & SemEval* & SEntFiN 1.0 \\
        \midrule
        \textbf{Time Stamp}      & \color{green}{Yes}     & \color{green}Yes    & \color{green}Yes    & \color{green}Yes   & \color{green}Yes   & \color{red}No & \color{red}No   & \color{red}No   & \color{red}No \\
        \textbf{Text Type}       & \color{green}Article     & \color{green}Article    & \color{green}Article    & \color{green}Article   & \color{green}Article   & \color{red}Sentence & \color{red}Sentence & \color{red}Headline & \color{red}Headline \\
        \textbf{Number of News}          & 15698563 & 8556324 & 3252885 & 447341 & 800974 & 1000  & 21867 & 1142 & 10753 \\
        \textbf{Symbol}          & \color{green}Yes      & \color{red}No      & \color{green}Yes     & \color{red}No     & \color{red}No     & \color{red}No    & \color{red}No    & \color{red}No   & \color{red}No \\
        \textbf{Summarization}   & \color{green}Yes      & \color{red}No      & \color{red}No      & \color{red}No     & \color{red}No     & \color{red}No    & \color{green}Yes   & \color{red}No   & \color{red}No \\
        \textbf{Sentiment Score} & Integer       & -     & -    & -    & -    & Integer     & -   & Real & Integer \\
        \textbf{URL}             & \color{green}Yes      & \color{red}No      & \color{green}Yes     & \color{red}No     & \color{red}No     & \color{red}No    & \color{red}No    & \color{red}No   & \color{red}No \\
        \textbf{Language}        & Many     & Eng     & Eng     & Eng    & Ru     & Eng   & Eng   & Eng  & Eng \\
        \textbf{Stock Price}     & \color{green}Yes      & \color{red}No      & \color{red}No      & \color{red}No     & \color{red}No     & \color{red}No    & \color{green}Yes   & \color{red}No   & \color{red}No \\
        \bottomrule
    \end{tabular}
    \captionsetup{font=small}
    \caption{Comparison of existing datasets for Time Series Financial Analysis. FNSPID stands out with the highest volume of news data and includes unique features not found in other benchmark datasets. In the label, SemEval* stands for SemEval-2017 Task5 dataset.}
    \vspace{-6mm}
    \label{tab:comparison}
\end{table*}

"Modern portfolio theory", by Harry Markowitz, emphasizes the market correlation~\cite{fabozzi2002legacy, konstantinov2020network}. Recent studies have highlighted a strong positive correlation of sentiment information, including news, blogs, and social media, with the stock market trends~\cite{darapaneni2022stock, hsu2021news}. The advent of advanced Large Language Models (LLMs) like ChatGPT and GPT-4 developed by OpenAI~\cite{openai-gpt-3} have significantly improved the accuracy of sentiment analysis in this context. The research from Lopez-Lira mentioned LLMs, like GPT-3, struggled with accurate market return forecasting. However, cutting-edge models, like GPT-4, achieved the highest Sharpe ratios, demonstrating increased reliability~\cite{lopezlira2023chatgpt}.

Beyond sentiment analysis by GPT-4, LLMs serve diverse roles in finance, including RL and specialized financial LLMs like FinGPT~\cite{wang2023fingpt} and FinRL~\cite{FinRL2024Liu}. Integrating numerical data into language models is challenging, but multi-modal models embedded stock prices and news data enhance accuracy~\cite{Sentiment2021Mohan, Gupta2020Multimedia}. However, this approach may not optimize general pre-trained LLMs due to potential information loss from using only sentiment scores. Meanwhile, the lack of comprehensive and integrated datasets has significantly limited advancing research, particularly in implementing more sophisticated models like those based on transformer technology, which could significantly enhance financial analysis.
To address this gap, previous datasets, such as Philips's news from Bloomberg and Reuters~\cite{BloombergReutersDataset2015Philippe}, and Yutkin's news from Lenta~\cite{Yutkin2019LentaRussian}, along with contributions from sources like Benzinga, have been valuable. However, these datasets often lack sufficient data volume for training large models and do not always include corresponding stock prices. Moreover, the available news data frequently lacks a structured time series format, posing challenges for sequence-to-sequence prediction models. To solve these issues, we introduce the Financial News and Stock Price Integration Dataset (FNSPID). This dataset uniquely combines time series news and stock prices, providing a groundbreaking resource for financial market analysis.

\textbf{1. Utilization of ML for Finance:} FNSPID is designed for stock market prediction ML model development. Its textual and numerical data integration enhances model functionality and provides a solid foundation. The dataset is not limited to ML but also other financial sentiment-price correlation analyses and offers nuanced insights into market dynamics and stock price trends.

\textbf{2. Insight from FNSPID:} Experiments utilizing FNSPID demonstrated larger datasets lead to better performance in price prediction; quality of sentiments leads to a positive effect on boosting accuracy.

\textbf{3. Apply and reproduce FNSPID:} FNSPID founded research in the financial domain for sentiment analysis, LLM fine-tuning, and research on DL models. We provided reproducible examples with instructions for expanding the datasets.

In Section 2 of this paper, we describe the related work for financial datasets. In Section 4, we show that \textit{FNSPID is a significant advancement in financial forecasting, filling key gaps in existing resources}. In Section 5, we present that \textit{FNSPID enables the training of larger stock prediction models more accurately in market dynamics analysis} with the advantage of a large amount of time series news with stock price data. The dataset's various features, including data attributes, enabled diverse applications beyond machine learning, encompassing sentiment analysis, trend evaluation, and risk assessment. Section 3 describes how FNSPID is constructed; Section 6 discusses FNSPID application and ethics. Our objective is to demonstrate that \textit{(a) FNSPID supports research in advanced ML techniques. (b) Beyond academia, FNSPID supports precise financial tools and aids in better capital allocation.}

\section{Related Work} 
\subsection{Evolution of Financial Analysis Models}
Financial analysis has undergone significant evolution, especially with the advent of sophisticated models. Key examples among these are the Fama-French Three-Factor Model (FFE)~\cite{fama1992cross}, and the Chen, Roll, and Ross Arbitrage Pricing Theory (APT) model~\cite{chen1983APT}. As shown in Appendix Section A.1, these models consider factors such as market risk, size, and value to understand asset prices. While they contribute to long-term analysis, they lack the granularity to forecast short-term price movements effectively, such as the exact peaks or troughs in stock prices. This limitation has prompted the exploration of additional data sources to enhance the predictive accuracy in financial analysis.

To enhance accuracy in time series financial analysis, common tools like Autoregressive Integrated Moving Average (ARIMA)~\cite{maciel2008design} and Generalized Autoregressive Conditional Heteroscedasticity (GARCH)~\cite{ARIMA2022Li} are widely implemented. These tools, along with traditional metrics and technical analyses, support market trend analysis but have limitations due to the subjectivity and bias in investors' decision-making. 

However, the emergence of machine learning (ML) has greatly improved accuracy in timing market entries and exits. The use of ML in financial markets has evolved from basic techniques like Linear Regression and SVM to advanced methods such as LSTM, RNN, and Deep Q-learning~\cite{RLstockprediction2001Jae, data4030110, WU2020142}. Direct application of reinforcement learning in stock trading strategies shows promise~\cite{DeepLearning2021Yang}, and DL models prove effective in stock market analysis. 

Recent studies highlighted ML, utilizing diverse data sources like real-time news, social media sentiment, and economic indicators, becomes a robust alternative to traditional stock prediction methods. Sheth and Kurani confirm this by comparing ML approaches to traditional models, emphasizing ML's ability to detect complex, non-linear patterns and adapt to changing market conditions~\cite{Sheth2023, Kurani2023}. LSTM models have also shown impressive error reduction (MAE), underscoring ML's continuous learning and updating capabilities~\cite{Billah2024}, making it a reliable tool in financial forecasting, which nicely moves beyond historical trends to incorporate and adapt to current market dynamics.

Financial news significantly influences the overall movement of the market, particularly evident by a GARCH model analysis during the 2008-2009 financial crisis by recent research~\cite{SAKARIYAHU2023101866}. In terms of sentiment analysis, numerous studies have highlighted the conditional efficacy of combining sentiment analysis with machine learning techniques to enhance prediction accuracy in various domains, including financial markets~\cite{lopezlira2023chatgpt, Shi2023Layout, Sinha2022, Venuti2021Predicting, Wang2017Exploring, Liapis2023Investigating, LiPan2020}. Specifically, Venuti~\cite{Venuti2021Predicting} employed a graph-based machine learning framework to analyze company relationships, although it did not incorporate real-time market data. Similarly, Wang et al.~\cite{Wang2017Exploring} focused on integrating sentiment analysis with machine learning for stock volatility prediction, but without engaging deeply with specific financial metrics. The impact of news sentiment on financial markets, as studied by Qudah and Rabhi, involved analyzing sentiment datasets but did not consider individual investor behaviors~\cite{QudahRabhi2016}. 

Remarkably, recent studies using pre-trained language models like GPT-3.5 for generating news articles and then applying sentiment analysis to predict stock prices have shown promising results, outperforming traditional sentiment analyzing algorithms and methods~\cite{lopezlira2023chatgpt, wang2023fingpt}. These advancements include FinGPT, a model trained on a large corpus of financial news for generating high-quality articles. Furthermore, Gupta (2020) delved into the correlation between sentiment data and stock prices, enhancing predictive models~\cite{Gupta2020Multimedia}. Recent research conducted by Zhou~\cite{zhou2023OnefitsAll} explores the capabilities of LLMs backbone models for time series prediction. This study demonstrates considerable prediction accuracy in utilizing these models, despite their frequent oversight for non-financial market applications and the challenges posed by the scarcity of robust datasets.

\subsection{Existing Stock Dataset} 
Many previous works have demonstrated that sentiment analysis's accuracy for various DL models is highly dependent on the amount of training data and the quality of the training data.~\cite{Luo2021Data, Abdelwahab2015Effect, Antonowicz2022DataQuality, Riyadh2022GANBElectra, Ibrahim2017Quality}. The financial dataset landscape is evolving, with a growing emphasis on integrating sentiment analysis and news content for more accurate stock market predictions. Lutz~\cite{Lutz2018} offers a dataset that provides binary sentence-level sentiment analysis, categorizing financial news as positive or negative, along with textual representations. However, this dataset does not include detailed company financials, presenting a unique perspective on financial news sentiment.In contrast, Farimani~\cite{Farimani2021} introduced a dataset that combines latent economic concepts, news sentiment, and technical indicators, where all the data is provided in time series which is very important for combining the sentiment information with stock price information, but the sentiment information included is the currency exchange rate and correlated news. Meanwhile, it falls short in terms of in-depth trading data. 
Cortis~\cite{SemEval2017} provided a dataset for fine-grained sentiment analysis of financial microblogs and news, including sentiment scores and lexical/semantic features. However, this dataset contains only a limited amount of news headlines (1142 articles) and employs a proprietary formula for sentiment scoring, which may not accurately reflect actual news sentiment. Moreover, Sinha et al~\cite{Sinha2022} SEntFiN 1.0 dataset, notable for its entity-sentiment annotations and extensive database of financial entities, provides relatively more handy information than the work provided previously. Nevertheless, it does not include the timestamp which plays a critical role in aligning sentiment data with price data. Meanwhile, short headlines were the only information provided, which can be inaccurate in determining the sentiment within short 
paragraphs and the small dataset does not provide enough information to support the sentiment information training. 

To address this, Philippe's dataset, sourced from Bloomberg and Reuters, offers a large collection of financial news time series for analysis~\cite{BloombergReutersDataset2015Philippe}. However, it lacks entities for target sentiment analysis with raw, unprocessed news content. This could impact forecasting accuracy. Recent innovations include a novel stock price prediction method combining numerical data with social media text features, using a deep reinforcement learning model, and introducing new dynamic datasets for evaluating prediction models~\cite{FinRL2024Liu}. Meanwhile, the Finnhub dataset provides stock prices with correlated news in a time series by calling the API. It is beneficial for sentiment analysis research, though the proprietary and the lack of sentiment analysis makes the model training inconvenient. However, the dataset is not tested on the quality from previous research~\cite {finnhub}.
    
FNSPID encompasses a wide range of financial news in English and Russian, covering 1999 to 2023. FNSPID correlates news with stock prices, serving as a valuable resource for sentiment analysis and stock price prediction. Concerning data quality, especially in the context of the proliferation of 'fake news,' our dataset exclusively sources information from trusted financial news platforms like NASDAQ. This ensures the reliability and relevance of the data for sentiment analysis and stock market prediction, setting a standard for dataset integrity in financial modeling research.
               
One of the challenges in this field is the limited access to high-quality, open-source datasets. For instance, datasets and models like Finchat and BloombergGPT, while valuable, often come with accessibility restrictions and are not openly available for academic research. This limitation hampers the ability of researchers to fully explore and develop innovative models in financial prediction. This work seeks to address this gap by providing a dataset that is both comprehensive and accessible, paving the way for more open and inclusive research in the field of financial modeling.
    
\begin{figure}[!t]
    \centering
    \includegraphics[width=1\linewidth]{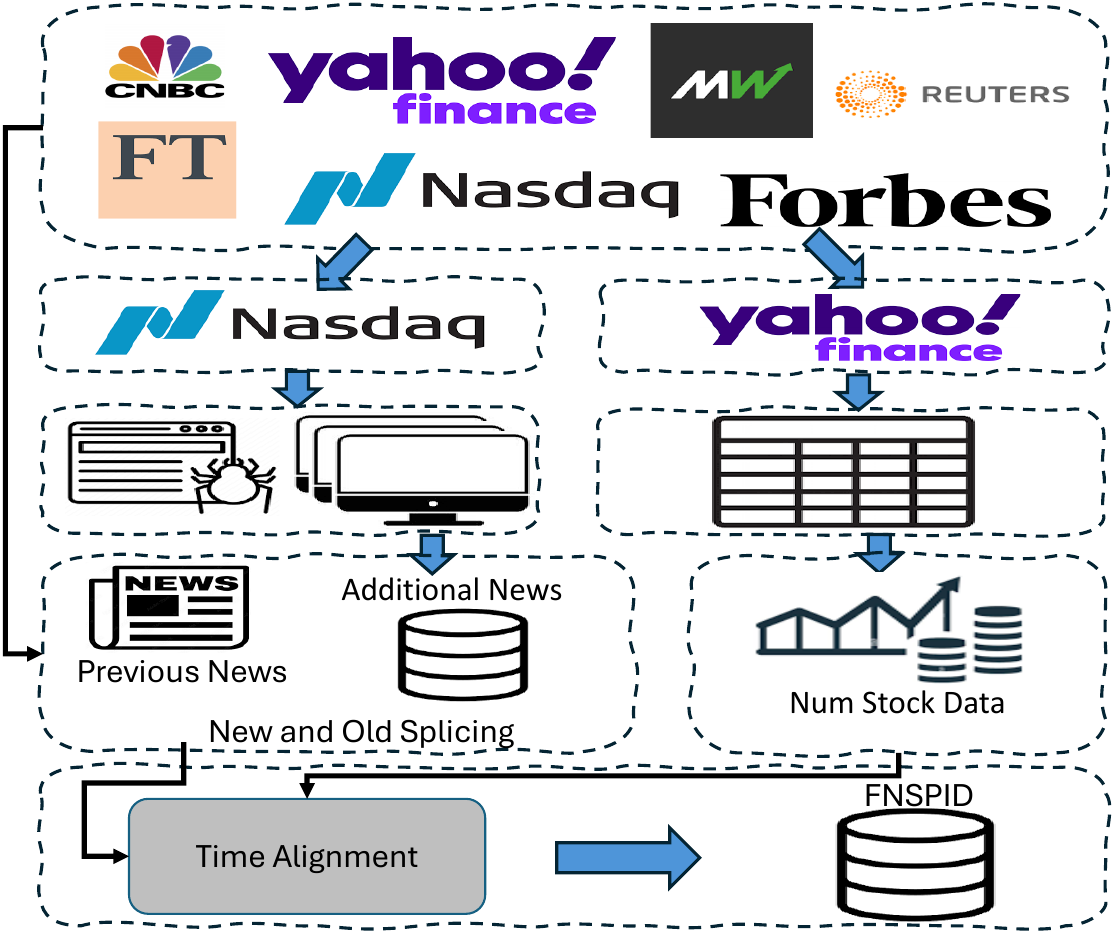}
    \vspace{-6mm}
    \captionsetup{font=small}
    \caption{Data Collection Process from website selection in the first level box; data segmentation in second level boxes; data collection for web scraping on left and numerical data collection on right; data organization on fourth level boxes and final FNSPID build-up on the last level box.}
   \vspace{-5mm}
    \label{fig:FNSPID_Collection_Procedure}
\end{figure}


\section{Constructing FNSPID} 
FNSPID is a carefully curated collection of numerical and sentiment data. In this section, we are going to describe the construction of the main part of FNSPID which includes all the sentiment and numerical information~\textbf{(Task 1)}. Next, we are going to describe how we build up the summarized sentiment dataset~\textbf{(Task 2)}. Lastly, we talked about how we built up the quantified sentiment dataset~\textbf{(Task 3)}.

\textbf{Data Sources:} As shown in Figure~\ref{fig:FNSPID_Collection_Procedure}, we obtained numerical stock data from Yahoo Finance's API and sentiment data from various reputable sources. Our exploration led us to numerous news websites such as Bloomberg, Yahoo Finance, Reuters, Forbes CNBC, etc. However, all of these websites have limited policies on data usage. We collected news from NASDAQ, which involves a two-stage process. Initially, we collected headlines and URLs from NASDAQ for each stock in the list by the Python package \textit{Selenium}. Then, we extracted news content from URLs to build the textual part of the dataset. To enhance data integrity and diversity and prevent website bias, we processed and combined previous raw data from Bloomberg, Reuters, Benzinga, and Lenta, which offer comprehensive or longer-retained information. Combining the two parts, we build up the FNSPID Task 1.

\textbf{Data Ethics:} In collecting data from NASDAQ, we rigorously adhered to ethical standards, consulting the robots.txt file to ensure compliance with website policies and avoiding potential conflicts of interest. Mindful of copyright and regional policies, we restricted our collection to content freely available without premium access or subscription requirements. Given the absence of an API, we resorted to web scraping to acquire the necessary news data. By acknowledging and confirming the license of previous work, we combined the existing processed data as part of FNSPID.

\begin{figure}[t]
    \centering
    \includegraphics[width=1\linewidth]{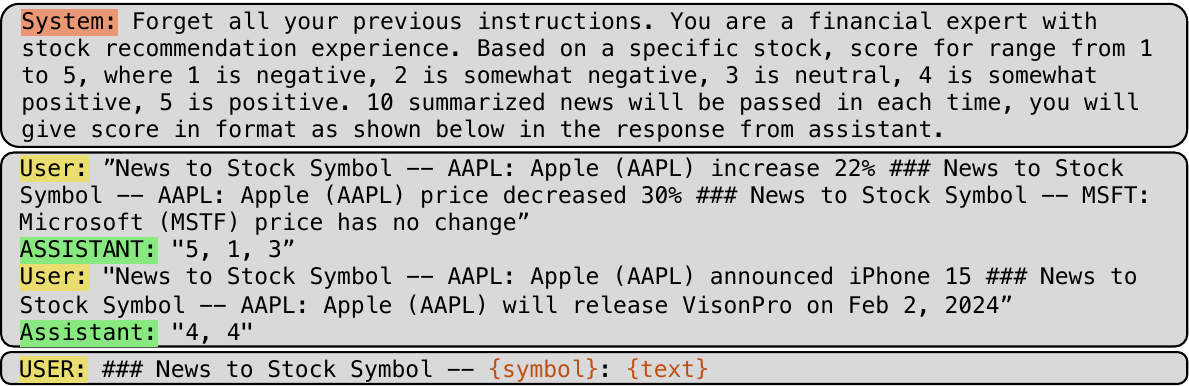}
    \vspace{-6mm}
    \captionsetup{font=small}
    \caption{Example ChatGPT Prompt: The first section is the system prompt, defining constraints and specifying the task for ChatGPT. In the second section, two examples are included to guide ChatGPT on the desired content for the response. Subsequently, the summarized news is fed into ChatGPT for sentiment score labeling. \textit{\{symbol\}} is the stock symbol variable input and \textit{\{text\}} is the news variable input.}
    \vspace{-4mm}
    \label{fig:ChatGPT_prompt}
\end{figure}

\begin{figure}[!t]
    \centering
    \includegraphics[width=1\linewidth]{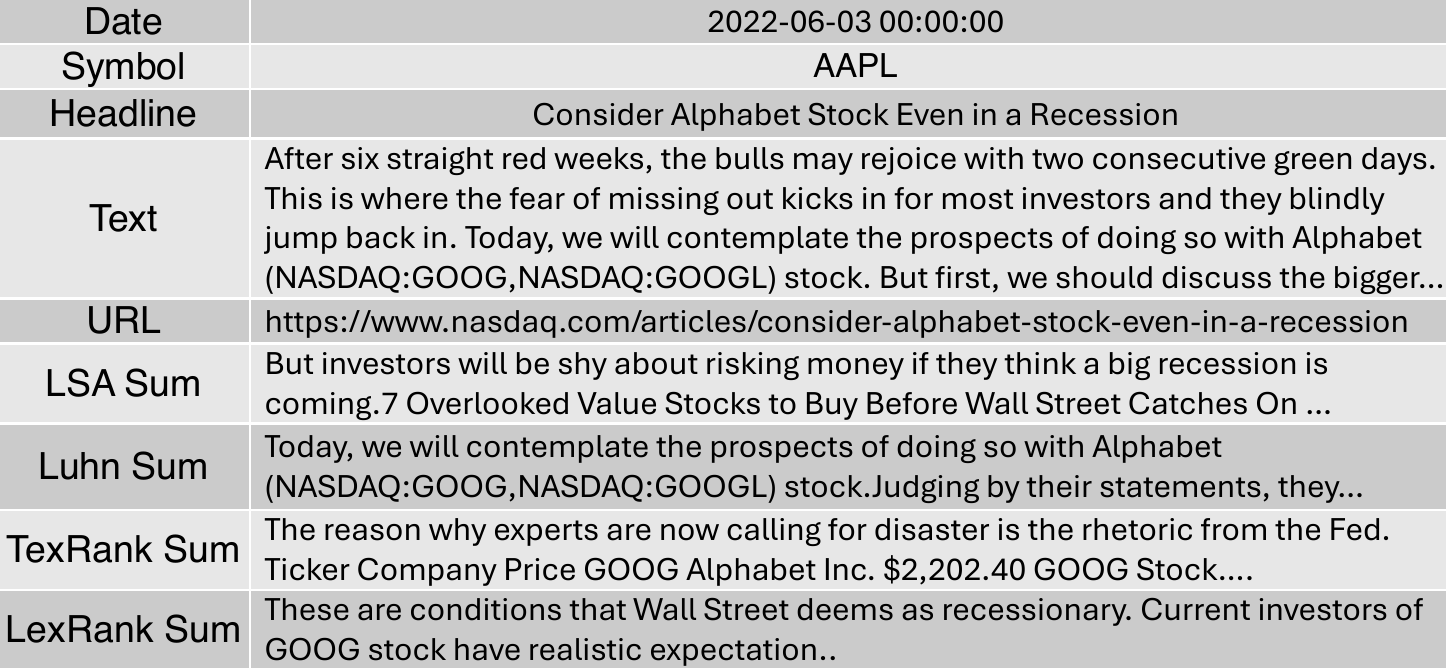}
    \vspace{-6mm}
    \captionsetup{font=small}
    \caption{Sentiment Data: Where 'Symbol' represents the stock code (e.g., AAPL for Apple Inc.); 'LSM Sum', 'Luhn Sum', 'TextRank Sum', and 'LexRank Sum' encapsulate the summarized news information generated by three different algorithms.}
   \vspace{-4mm}
    \label{fig:Sentiment_Example_data}
\end{figure}

\subsection{Data mining and processing}
After collecting the raw dataset containing numerical prices, URLs, news headlines, and news text, we performed extensive sentiment analysis by summarizing each article using four methods: LexRank, Luhn, Latent Semantic Analysis (LSA), and TextRank. Each method comes from the Python package Sumy, known for its robust summarization capabilities and rule-based tokenization approach. These summaries are crucial for handling token limitations and practical constraints in sentiment analysis \textbf{(Task 2)}. To enhance the summaries' relevance to the related stock, we introduced a weight model \(W_f\) detailed in Appendix A.2 to enhance the summary by giving more attention to the related stock. After sample reviewing, we set the summary length to 3 sentences to ensure the summaries concisely contained useful information, which is approximately one-eighth of the original length to keep the conciseness while avoiding losing the specificity. This step significantly reduced token usage for subsequent large language model analyses while reaching the critical point for ChatGPT's prompt stability giving out a stable answer. With this, we finished constructing the FNSPID Task 2.

\begin{figure}[t]
    \centering
    \includegraphics[width=0.8\linewidth]{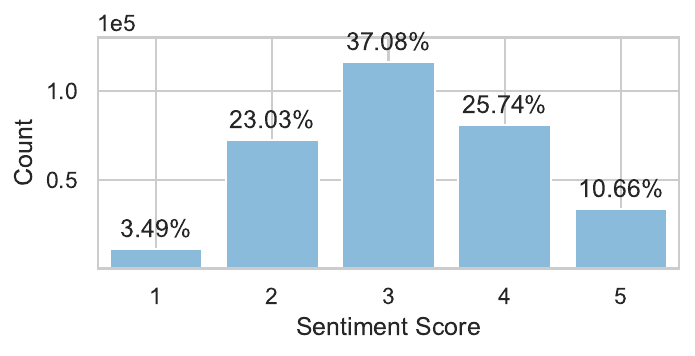}
    \vspace{-6mm}
    \captionsetup{font=small}
    \caption{Sentiment Distribution: 1 is negative, 2 is somewhat negative, 3 is neutral, 4 is somewhat positive, 5 is positive}
    \vspace{-6mm}
    \label{fig:Sentiment-distribution}
\end{figure}

\textbf{Sentiment Quantification} Neither early-state LLMs, like GPT-2 and GPT-3, nor time-series deep learning models can understand natural language properly. The limitation of computational resources does not allow most of the experiments including models like ChatGPT, which has hundreds of billions of parameters. However, previous work shows DL models could handle the sentiment signals properly~\cite{Luo2021Data, Abdelwahab2015Effect, Antonowicz2022DataQuality, Riyadh2022GANBElectra, Ibrahim2017Quality}. To meet the requirement, we incorporated a small dataset of news articles collected from 50 prominent US stocks from S\&P 500 with sentiment labels (Task 3). To integrate sentiment labels into the input without intensive human labeling, we utilized ChatGPT for sentiment analysis, acknowledging the challenges faced by conventional algorithms and language models, including GPT-2 and GPT-3, in accurately scoring sentiments~\cite{lopezlira2023chatgpt, fatouros2023transforming, kocon2023chatgpt, zhang2023instruct}. We opted for the output from the previous step from the LSA summarizer algorithm, which condensed the news content and provided ChatGPT with succinct yet comprehensive inputs for sentiment analysis. Figure~\ref{fig:ChatGPT_prompt} illustrates how the summarized news content is used as a user prompt for ChatGPT. To ensure effectiveness in maintaining the model’s stability, we input up to 10 data entries at a time into ChatGPT with a temperature setting of 0. During the experiment, the sentiment score from 1 to 5 is more stable than other distributions like -1 to 1, and 1 to 10. Among these, using decimals to describe sentiment distribution causes the most inaccurate sentiment representation. We employed a sentiment scoring scale that spanned from 1 to 5, as shown in Figure~\ref{fig:ChatGPT_prompt}, where 1 represented a negative sentiment, 2 was somewhat negative, 3 was neutral, 4 somewhat positive, and 5 was positive. This gradation in scoring facilitated a more detailed and subtle interpretation of news sentiment. In integrating these sentiment scores with other features of our model, we applied a consistent normalization approach. This was crucial to ensure that while the sentiment scores contributed to the model's training, they did not disproportionately influence its outcomes. From the sentiment score quantified by ChatGPT, we see an approximate of the normal distribution as shown in Figure~\ref{fig:Sentiment-distribution}.
\begin{equation}
S_{\left(t\right)}\:=\:3+\left(S_{\left(0\right)}-3\right)\cdot \:\:e^{-\lambda\:\cdot \:t} \\
\end{equation}

\textbf{Handling Data Gaps:} To address data gaps on dates without news information, we implemented an exponential decay method as shown in Equation 6, where \(S_{\left(t\right)}\:\) is the sentiment score after t days of the previous date that has news, \(S_{\left(0\right)}\) is the sentiment score of the day 0, \(\lambda\) is the decay factor which we choose \(\lambda=0.03\) and  \(t\) is the time after the first day of sentiment score been decided. \(3\) in the formula sets the decay target to the neutral value in the sentiment score. This method was used to extrapolate missing sentiment factors from the previous day's news, ensuring temporal continuity in our dataset. Additionally, for days with multiple news articles, we calculated the average sentiment score. This average represents the day’s overall sentiment, allowing for a more nuanced and accurate reflection of the day's market sentiment. Our decision to use averaging is based on the rationale that it provides a balanced representation of the day's sentiment, mitigating the influence of any single news item.

\begin{table}[t]
    \centering
    \small
    \begin{tabular}{|p{2.2cm}|p{0.6cm}|p{0.6cm}|p{0.6cm}|p{0.6cm}|p{0.6cm}|p{0.9cm}|}
        \hline
         Date & Open & High & Low & Close & Adj. & Volume \\
        \hline
        2023-12-28 00:00:00  & 194.14 & 194.66 & 193.17 & 193.58 & 193.58 & 34014500 \\
        \hline
        2023-12-27 00:00:00 & 192.49 & 193.50 & 191.09 & 193.15 & 193.15 & 48087700 \\
        \hline
        2023-12-26 00:00:00 &	193.61 &	193.89 &	192.83	& 193.05 &	193.05	& 28919300	\\
        \hline
        ... & ... & ... & ... & ... & ... & ... \\
        \hline
    \end{tabular}
    \captionsetup{font=small}
    \caption{Stock Numerical Data: 'Open' represents the opening stock price, 'High' indicates the highest price within the day, 'Low' signifies the lowest price within the day, 'Adj Close' represents the close price adjusted for dividends, and 'Volume' denotes the number of shares traded.} 
    \vspace{-6mm}
    \label{StockPrice_Example_data}
\end{table}

\section{FNSPID Property} 
Upon completion of the data mining and processing, the FNSPID is now primed for analytical examination. This section delineates the principal findings from diverse analytical approaches.

\textbf{Dataset Overview:} The FNSPID is comprehensive and varied, encompassing over 30 GB of data. As depicted in Table~\ref{StockPrice_Example_data}, we illustrate a sample of the time-series numerical price data included in our dataset. Figure~\ref{fig:Sentiment_Example_data} offers a glimpse into the sentiment data, encompassing URLs, news headlines, news text, sentiment scores, and articles summarized through four distinct methodologies. This diverse array of data points underscores the dataset's depth and breadth. The collective effort, requiring approximately 4TB of computing power and 45 days, reflects our commitment to overcoming these challenges and ensuring the robustness of our analysis.

Beyond the summarization, we expanded our analysis to include 50 stock samples selected from the top 50 influential stocks in the S\&P 500 as of 2024. These samples were incorporated into our batch for sentiment labeling, resulting in a total of 402,546 news items with assigned sentiment scores. 

    
    \subsection{Evaluation} 
    \begin{figure}[!t]
    \centering
    \includegraphics[width=1\linewidth]{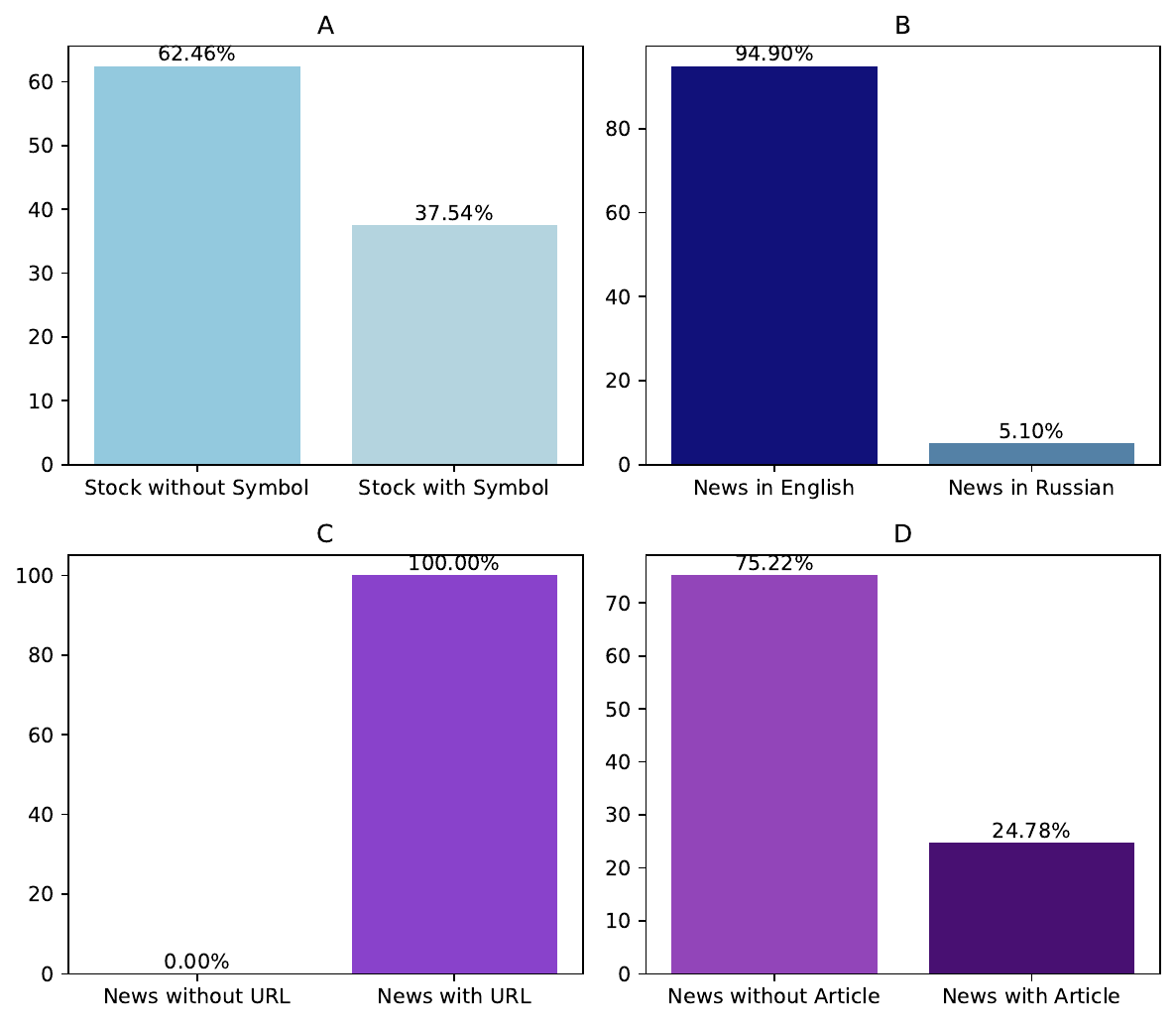}
    \vspace{-6mm}
    \captionsetup{font=small}
    \caption{Statistical Overview: In A, we provide information on news articles that include the stock symbol. The B displays the language distribution, encompassing English and Russian. In C, a comparison of the included URLs is presented. Finally, in the D, details are provided on the news text already incorporated in the dataset, along with potential expansions into additional text data.}
    \vspace{-4mm}
    \label{fig:feature_comparison}
    \end{figure}


    \textbf{Language Distribution:} Delving into the linguistic makeup of our dataset, we analyzed the percentage distribution of languages, notably Russian and English. This exploration provided critical insights into the multilingual nature of our data, as detailed in Figure~\ref{fig:feature_comparison}. Understanding this distribution is essential, as it reflects the global applicability and versatility of FNSPID.

    \textbf{News Article Segmentation:} We differentiated between news articles containing stock symbols and those without. This distinction is pivotal, revealing the extent to which stock-related news pervades our dataset. Figure~\ref{fig:feature_comparison} visually delineates this segmentation, offering insights into the dataset's alignment with stock market information.

    \begin{figure}[!t]
    \centering
    \includegraphics[width=1\linewidth]{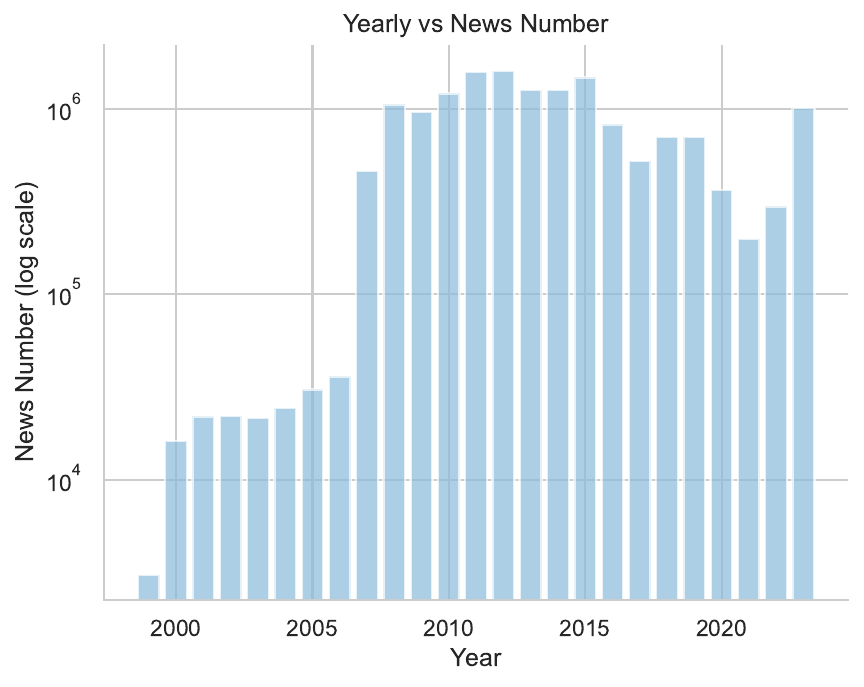}
    \vspace{-8mm}
    \captionsetup{font=small}
    \caption{News Count Over Time: This graph illustrates the number of news articles over the years, providing a comprehensive view of the distribution from 1999 to 2023.}
    \vspace{-6mm}
    \label{fig:Number_of_News_vs_Time}
    \end{figure}
    
    \textbf{Temporal Distribution Analysis:} Our exploration extended to the temporal distribution of news articles to discern trends and patterns over time. Figure~\ref{fig:Number_of_News_vs_Time} illustrates the volume of news articles per year from 1999 to 2023. This temporal analysis enriches our understanding of the dataset's evolution and the fluctuating dynamics of financial news coverage, offering a valuable perspective on the historical trends in financial reporting.

    Through these analyses, FNSPID emerges as a uniquely comprehensive and multi-faceted dataset, poised to facilitate advanced research in financial sentiment analysis and time-series prediction. The dataset's vast scope, multilingual capacity, and temporal depth make it an invaluable resource for researchers and practitioners in financial modeling and analysis.

\section{Experiement} 

    \begin{figure}[t]
        \centering
        \includegraphics[width=1\linewidth]{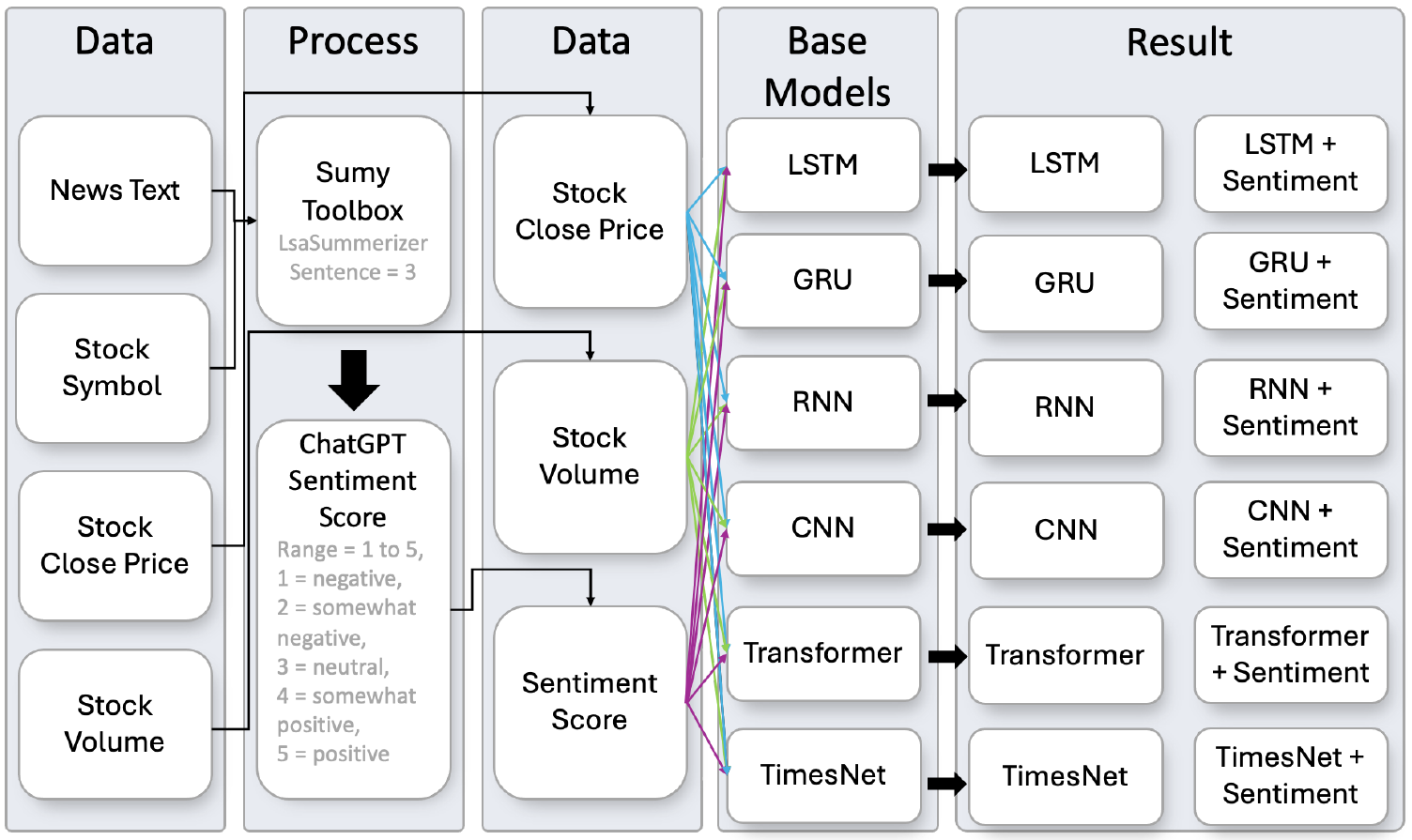}
        \vspace{-6mm}
        \captionsetup{font=small}
        \caption{Experiment Procedure: The experimental setup involves utilizing news text, stock symbol, stock close price, stock open price, and stock volume as inputs to predict the stock's close price. The news text is processed through the Lsa-summarizer, followed by ChatGPT sentiment quantification. The obtained sentiment score, stock close price, open price, and volume are input into CNN, RNN, LSTM, GRUs, Transformer, and TimesNet. Concurrently, a reference group is established, incorporating only stock open price, close price and volume as input variables.}
        \vspace{-5mm}
        \label{fig:Experiment_Procedure}
    \end{figure}

    \begin{table*}[!t]
    \centering
    \begin{tabular}{c|p{1.8cm}|ccc|ccc|ccc|ccc}
        \toprule
         & \textbf{Dataset} & A-Sen. & A-Sen. & A-Sen. & A-Non. & A-Non. & A-Non. & B-Sen. & B-Sen. & B-Sen. & B-Non. & B-Non. & B-Non.\\
        \midrule
        \# & \textbf{Name} & MAE & MSE & R$^2$ & MAE & MSE & R$^2$ & MAE & MSE & R$^2$  & MAE & MSE & R$^2$ \\
        
    \midrule
    5 & LSTM & .02599 & .00157 & .87115 &
    .02530 & .00148 & \textbf{.88016} & .02677 & .00160 & \textbf{.86811} &
    .02523 & .00142 & \textbf{.88181} \\
    & CNN & .06180 & .00712 & .48205 &
    .04913 & .00475 & .61811 & .04236 & .00354 & .71668 &
    .04522 & .00398 & .66687 \\
    & GRU & .02474 & .00143 & .88588 &
    .02494 & .00141 & .88302 & .02631 & .00154 & .86756 &
    .02470 & .00139 & .87746 \\
    & RNN & .04152 & .00355 & .72957 &
    .03353 & .00251 & .81128 & .04315 & .00339 & .54265 &
    .03898 & .00291 & .65470 \\
    & \textbf{Transformer} & \textbf{.01801} & \textbf{.00058} & \textbf{.87260} & \textbf{.01883} & \textbf{.00060} & .86659 & \textbf{.01700} & \textbf{.00060} & .84659 & \textbf{.01007} & \textbf{.00021} & .94629 \\
    & TimesNet & .02847 & .00148 & .63407 &
    .02225 & .00089 & .81824 & .03441 & .00194 & .51742 &
    .02697 & .00129 & .69189 \\
    
    \midrule
    25 & LSTM & .02569 & .00155 & .87040 &
    .02482 & .00141 & .87627 & .02569 & .00146 & .86889 &
    .02706 & .00178 & .86401 \\
    & CNN & .04520 & .00402 & .69021 &
    .04271 & .00371 & .71418 & .04201 & .00365 & .71958 &
    .04161 & .00354 & .72290 \\
    & GRU & .02696 & .00178 & .86873 &
    .02484 & .00145 & .88233 & .02848 & .00192 & .86129 &
    .02523 & .00142 & .87175 \\
    & RNN & .03829 & .00311 & .73611 &
    .03426 & .00277 & .76536 & .03828 & .00293 & .68064 &
    .03975 & .00280 & .58985 \\
    & \textbf{Transformer} & \textbf{.00757} & \textbf{.00008} & \textbf{.98304} & \textbf{.00711} & \textbf{.00008} & \textbf{.98178} & \textbf{.00943} & \textbf{.00013} & \textbf{.96811} & \textbf{.00763} & \textbf{.00009} & \textbf{.97948} \\
    & TimesNet & .02347 & .00093 & .79670 &
    .02364 & .00093 & .77555 & .02412 & .00104 & .77040 &
    .02319 & .00091 & .78261 \\

        \midrule
    50 & LSTM & .02493 & .00170 & .85585 &
    .02510 & .00145 & .87988 & .02772 & .00168 & .83983 &
    .02590 & .00154 & .86678 \\
    & CNN & .03550 & .00289 & .73355 &
    .04126 & .00343 & .73344 & .04092 & .00346 & .74825 &
    .04129 & .00343 & .73457 \\
    & GRU & .02769 & .00209 & .82767 &
    .02612 & .00166 & .87071 & .02671 & .00160 & .85643 &
    .02587 & .00150 & .86944 \\
    & RNN & .04154 & .00389 & .61744 &
    .03343 & .00243 & .78635 & .03849 & .00317 & .75238 &
    .03658 & .00289 & .74494 \\
    & \textbf{Transformer} & \textbf{.00544} & \textbf{.00005} & \textbf{.98785} & \textbf{.00615} & \textbf{.00006} & \textbf{.98592} & \textbf{.00488} & \textbf{.00004} & \textbf{.99109} & \textbf{.00614} & \textbf{.00007} & \textbf{.98527} \\
    & TimesNet & .02577 & .00106 & .73819 &
    .02181 & .00084 & .80573 & .02119 & .00077 & .82663 &
    .02551 & .00118 & .72460 \\
        \bottomrule
    \end{tabular}
    \captionsetup{font=small}
    \caption{Experiment Evaluation via 50 epochs of training, A-Sen. is ChatGPT labeled sentiment dataset result, B-Sen. is the TextBlob labeled sentiment dataset, A-Non., and B-Non. are the numerical data only dataset for experiments A and B. \# is the number of stocks used in training for 5,25,50.}
    \vspace{-4mm}
    \label{Experiment_result}
    \end{table*}    

To validate the FNSPID, we not only analyzed it statistically but also conducted experiments to test its reliability. In this section, we use the quantity and quality tests to examine the dataset's overall performance. This section outlines our experimental strategy, showcasing the dataset’s robustness in real-world applications.

\subsection{Quantity Test} 
For stock price predictions, people use numerical data and sentiment information as inputs to predict the short-term stock market behavior. Different models recognize different data patterns which leads to variations in the prediction performance. We used the FNSPID Task 3 to conduct experimental analysis, aiming to research the effectiveness of the quantity of news in the models. As shown in Figure~\ref{fig:Experiment_Procedure}, we conducted a comparison of DL methods in stock price prediction. The choice of LSTM, RNN, Convolutional Neural Networks (CNN)~\cite{chen2016financial}, and Gated Recurrent Units (GRU)~\cite{shen2018deep} as our primary models for validating the traditional method's performance of FNSPID. Beyond that, we also experimented with more novel methods in financial predictions: 4-layers Vanilla Transformer~\cite{vaswani2017attention} and 4-layers TimesNet~\cite{wu2023timesnet}~\cite{shen2018deep} which are both proficient in time series prediction. The normalizer models are placed in Appendix A.3. During the qualitative experiments, input features include open price, close price, and trading volume as baseline input features. We used 50 days of information and predicted 3 days in the future. Experiments were conducted for training with different numbers of stocks: 5 stocks (n = 11277), 25 stocks (n = 43192), and 50 stocks (n = 127937). 100 epochs was used for each training set. After the model training we used 5 stocks for evaluation, among them, we eliminated one outlier from the experiment result and gave the average value as the result.

\textbf{Test Results:} The result of the quantitative analysis for FNSPID is shown in Table~\ref{Experiment_result}, Part A-Sen. and A-Non. where the A-Sen. represent the experiment A with sentiment input, and A-Non. represent the experiment A excluded the sentiment information. The experimental result demonstrated on average 6.29 percent improvement of \(R^2\) from 5 stocks of training to 25 stocks of training among all 6 models we conducted.
The Transformer based model has the highest accuracy \(R^2 = 0.988\),the LSTM in second place achieved an accuracy of \(R^2 = 0.856\), and GRU model in third place got \(R^2 = 0.827\). Meanwhile, the RNN model had the worst performance \(R^2 = 0.617\).
Noticeably, Transformer overall has the best performance on accuracy in general which achieved \(R^2 = 0.988\) for accuracy where the second place is LSTM(\(R^2 = 0.856\) ) which is more than 0.13 \(\Delta R^2\) in difference with Transformer model. 
Through these meticulous experiments, we demonstrated the practical application and robustness of the FNSPID dataset, underscoring its value in financial modeling and sentiment analysis research. In general, in the trend analysis, a larger training dataset can lead to better performance of the financial stock prediction, which is a limitation of small datasets.

\subsection{Quality Test} 
With the sample model parameters the same as in the experiment for quantitative experiments, we compare the different models' training performance based on the sentiment from FNSPID Task 3 and the Experimental dataset explicated from FNSPID by using the TextBlob labeled information. The FNSPID Dataset Task 2 is ChatGPT labeled information. The Textblob sentiment information represents the combination of mathematical algorithms and small NLP models in sentiment score labeling.

From the experiment, The FNSPID Dataset Task 2 in Table~\ref{Experiment_result} Part A has a positive effect on the improvement in accuracy. Where the Textblob sentiment in Table~\ref{Experiment_result} Part B, hurts model training.

To avoid the initial randomness of the model, which has a significant impact, we conducted 5 tests to evaluate the results and calculate their average values for the experiment results. The experiment showed Sentiment quality and data quality affect the overall performance of the data when implementing the dataset in financial forecasting DL model training. In comparing (Transformer) sentiment and non-sentiment, while FNSPID Task 3 has a 0.2\% improvement, the Textblob sentiment has a -1.16\% impact on the overall stock price prediction.  

\textbf{Sentiment effectiveness:} After repetitions of experiments, in Table~\ref{Experiment_result}, we find only the transformer model has a positive effect on the improvement with including sentiment information, while TimesNet occasionally has a positive effect. We conclude other models do not have a very nice comprehension of when we integrate the sentiment information into the model and take the sentiment information as the noise.  It is also noticeable, that in small dataset training (when only 5 news), the LSTM outperforms the Transformer in training, while as the dataset goes larger, the Transformer has significant improvement in the accuracy of prediction.

\textbf{Discussion:} Models' hyper-parameters fine tuning can change the performance. However, to compare the models, we have to set the model parameters as close as possible, which could potentially hurt the performance of individual models.  We use the sentiment scoring on a scale of 5 to represent the sentiment information. The sentiment labeling methods could lead to some of the information from paragraphs being lost and cause the under performance of sentiment information in stock price prediction. Previous research has shown that financial news significantly impacts stock prices~\cite{allen2019daily}. However, our experiment revealed only a minor improvement in model performance, attributable to two main factors: firstly, the models' already high prediction accuracy makes further improvements challenging; secondly, potential delays in news dissemination may delay its impact on stock prices.

\textbf{In conclusion}, we summarize 3 points from the experiment based on FNSPID: 
\textbf{1.} Both the quality and quantity of the dataset largely affect the stock price prediction. 
\textbf{2.} High-quality sentiment information has a positive effect on transformer-based training.
\textbf{3.} The transformer-based model surpasses traditional time series models and novel methods like TimesNet in stock price prediction.

\section{FNSPID application and ethics} 
This discussion delves into the intricate interplay between machine learning methodologies and financial market analysis, as evidenced in our dataset. We critically examine the potential applications and ethical facets associated with our research.

\subsection{Challenges on FNSPID Construction}
In our data mining endeavor, after extracting data from Nasdaq, we experimented with various sentiment analysis methods. Tools like NLTK and TextBlob, alongside compact machine learning models, showed promise in interpreting simple sentiments, as in the phrases 'I hate you' and 'I love you.' However, their efficacy waned when tasked with parsing complex paragraphs from financial news sources. Larger models, including BERT, also fell short in yielding accurate sentiment predictions, as corroborated by Lopezlira et al. (2023)~\cite{lopezlira2023chatgpt}. These limitations led us to exclude sentiment scores from our final analysis. The cost and practicality constraints further sidelined the use of advanced tools like ChatGPT. Nevertheless, we provide code in subsequent sections for users interested in calling APIs for sentiment analysis and tailored to their specific needs.

Our preliminary trials with ChatGPT for sentiment scoring underscored challenges in achieving consistent outputs. Despite uniform instruction prompts, the variability in results pointed to a need for enhanced stability and interpretative precision in ChatGPT, particularly for diverse and complex financial texts. As mentioned in the data mining section, as the summarized sentence gets longer and is used as the user prompt input, the stability of the ChatGPT model still needs more development. When summarized texts contain more than 3 sentences, the model becomes less stable. Beyond that, long text input from the news given to ChatGPT will distract the model from giving the correct sentiment score.

\subsection{FNSPID Applications} 
This section discusses the great potential of implementing FNSPID in financial prediction and other aspects. We hope the quality, quantity, and diverse applications of FNSPID offer unparalleled opportunities to researchers in financial market analysis and beyond. 


\textbf{Multimodal models training:} Developing a dataset that merges textual and numerical inputs is crucial for creating multi-modal models, particularly in time series stock market prediction. Such a dataset could improve model robustness by leveraging the synergy between different data types. In addition to that, the current reliance on sequential data in reinforcement learning (RL) can be augmented by integrating a correlated dataset~\cite{wang2023fingpt}. This approach could significantly strengthen RL algorithms, especially in predicting stock market trends. For small and fast-deployed models that cannot understand natural languages, the FNSPID Task 3 enables the training.

\textbf{Sentiment Data in Market Prediction:} Evaluating the impact of sentiment data on market prices can draw insights from Modern Portfolio Theory. Parallel processing of news for multiple stocks could refine market predictions and reinforce RL algorithms.

\textbf{Correlation Analysis:} The dataset is pivotal in analyzing the correlation between sentiment information and stock prices, thereby enriching our understanding of market dynamics. FNSPID provides aligned sentiment-numerical data, which enables more accurate sentiment labeling, which is very important in quantitative analysis for investment banking. Beyond that, the FNSPID can be used for anomaly detection by recognizing the pattern of news that happened before the greater recession and helping with financial risk management and abnormal movement forecasting.

\textbf{Financial Generative AI:} Given the advantage of the quantity in the FNSPID, this dataset can aid in refining LLMs for improved financial advisory performance, leading to the development of advanced AI financial assistants.

\vspace{-2mm}
\subsection{Dataset Ethics} 

In reaffirming our commitment to ethical data collection practices, we meticulously adhere to a broad spectrum of ethical considerations that extend beyond the scope of website policies during web scraping. Our vigilant and multifaceted approach to ethics particularly focuses on privacy concerns in financial data analysis and the potential misuse of predictive models, ensuring our research practices meet the highest ethical standards.

\textbf{Privacy Concerns in Financial Data Analysis:} Financial data is inherently sensitive and requires robust protocols to ensure privacy and data security. Our methodology includes the implementation of advanced anonymization techniques to protect personal identifiers, and our data handling processes comply with international data protection regulations like GDPR and CCPA, ensuring the utmost privacy and confidentiality.

\textbf{Potential Misuse of Predictive Models:} Predictive models in financial contexts offer significant insights but also carry risks of misuse or unintended consequences. We have conducted an extensive ethical review of our predictive algorithms, incorporating fairness audits to prevent biases and ensure these models do not enable discriminatory practices. Clear guidelines for model usage have been established, preventing their application in ethically questionable contexts.

\textbf{Transparency and Data Marking:} Upholding our ethical commitment, every data point in our dataset is transparently marked and rigorously referenced. This practice not only bolsters our research's credibility but also promotes accountability and reproducibility within the academic community.

In conclusion, through these comprehensive measures, our study not only adheres to but also advances the discourse on ethical considerations in financial data analysis. We remain steadfast in our commitment to responsible and ethical academic inquiry, continuously striving to set exemplary standards in research ethics.


\section{Limitations and Future work} 
    \subsection{Limitations}

    Our dataset, while providing valuable insights, is not without limitations. The dynamic nature of website policies introduces a potential constraint, as future changes could impact the accessibility of our dataset. Maintaining adherence to current policies, any alterations in the future may necessitate adjustments to the data collection process. Additionally, the need for ongoing model validations remains crucial. As the field of stock market prediction evolves, continuous testing and validation of models with new datasets are imperative to assess adaptability and performance.
    
    \subsection{Future work}
    \textbf{Expand FNSPID:} Some of the existing stock news data are very popular and a large amount of the dataset has been included in a short period where the total amount of the dataset that could be collected is limited, which causes the news sentiment only ~20\% of the stock price data in timestamp value alignment. In the future we plan to expand the news dataset by developing an automated system that will keep the dataset up to date.
    \textbf{Exploring FNSPID:}
    FNSPID is one of the most complete datasets in aligning stock price and sentiment information. This dataset can help with the completion of a lot of new potential tasks. One is to construct multi-modal models based on the diverse data types within our dataset. This dataset is not only limited to the ML domain. It also provides the potential to analyze the influence factor of sentiment information on stock price fluctuation. The dataset will also promote news sentiment algorithms development. Beyond that, FNSPID can be used for stock correlation analysis as well. We hope FNSPID will become the main resource for a wide aspect of financial-related research.

By recognizing these limitations and proposing avenues for future exploration, we aim to encourage ongoing research efforts that build upon and refine the contributions of our dataset.

\begin{acks}  
We would like to express our gratitude to Dr. Dongkuan Xu, and Mr. Simon Anton from North Carolina State University for their 
 support and the reviewers for their comments. Additionally, we are thankful for Qin Yue from the University of Southern California's participation in this program.

\end{acks}
 \newpage
\bibliographystyle{ACM-Reference-Format}
\bibliography{reference}

 \newpage
\appendix

\section{Section A}

\subsection{Asset models}
The FFE model, presented in Equation 1, is building upon the Capital Asset Pricing Model (CAPM)~\cite{sharpe1964capital} which calculates the Asset Price (\(R_{it} - R_{ft}\)) at time \(t\) by considering three main factors: the excess return on the market portfolio index (\(R_{mt} - R_{ft}\)), the size premium (small minus big) (\(SMB_t\)), and the value premium (\(HML_t\)). This model brought a new perspective in understanding asset prices by integrating market risk, size, and value factors.
\begin{equation}
R_{it} - R_{ft} = \alpha_{it} + \beta_1 (R_{mt} - R_{ft}) + \beta_2 SMB_t + \beta_3 HML_t + \epsilon
\end{equation}
The APT model, represented in Equation 2, posits that the return of a portfolio \(r_{it}\) can be explained by multiple risk factors. These factors include changes in industrial production (\(IP_t\)), expected inflation (\(EI_t\)), unexpected inflation (\(UI_t\)), the excess return of long-term corporate bonds over government bonds (\(CG_t\)), and the excess return of long-term government bonds over T-bills (\(GB_t\)). The APT model provided a multi-factorial framework to asset pricing, acknowledging the influence of various macroeconomic factors.
\begin{equation}
r_{it} = \alpha_{it} + \beta_{iIP}IP_t + \beta_{iEI} EI_t + \beta_{iCG} CG_t + \beta_{iGB} GB_t + e_{it}
\end{equation}

\subsection{Summarize Algorithm}
To make the summarization more effective and include more attention toward the related stock, we introduced a weight model \(W_f\). In the package \textbf{sumy}, all the sentence summarizations are included, which means all the terms are chosen from the original sentence. On the other hand, exclusiveness means the sentence will be summarized in a new sentence instead of just picking out from the original sentence. In Equation (4), we first parse the paragraph \(T\) into individual sentences and we give a sentence weight \(W_s\) for weight \(m\) where in our method, we assign \(m=1\) to the sentence that contains the stock symbol. We also give the selected summarized sentences \(S_{sum}\) from Equation (5) a score for \(n\), where in our experiment we assigned \(n=1\), if the sentence is in the longer sentence \(S_{long}\). Finally, by implementing Equation (6), we add up the sentence weight \(W_s\) and summarized weight \(W_t\) to get the final weight score \(W_t\). For all other sentences, the weight will be 0. Finally, we sort the dictionary of the sentence set by weight and generate the final summarized sentence. With these measures, our dataset is well-prepared for detailed analysis.
\begin{equation}
    W_S(S, s) = \begin{cases}
  m & \text{if } S \in T \\
  0 & \text{otherwise}\end{cases}
\end{equation}
\begin{equation}
    W_t(S_{\text{sum}}, S_{\text{long}}) = \begin{cases}
  n & \text{if } S_{\text{sum}} \in S_{\text{long}} \\
  0 & \text{otherwise}\end{cases}
\end{equation}
\begin{equation}
 W_f = W_S + W_t
\end{equation}

\subsection{ML models Normalizer}
\begin{equation}
    S_n\:=\:\frac{V_n}{V_0}-1
\end{equation}
\begin{equation}
X_{scaled\:}=\:\frac{x\:-\:x_{min}}{x_{max\:-\:x_{min}}}
\end{equation}
Equation (7) represents the calculation of the normalized change in a value (\(S_n\)) relative to its initial value \((V_0)\) using the formula \(S_n = (V_n / V_0) - 1\). This equation measures how much the nth value has changed compared to the initial value, expressed as a fraction of the initial value.

Equation (8) is used for scaling a variable \((X_{scaled})\) to a range between 0 and 1. It rescales the variable x by subtracting the minimum value \((x_min)\) and dividing it by the difference between the maximum value \((x_{max})\) and the minimum value \((x_{min})\). This normalization process allows data to be represented within a consistent range, making it easier to compare and analyze.

\end{document}